# Interaction of Rydberg Excitons in Cuprous Oxide with Phonons and Photons: Optical Linewidth and Polariton Effect

**Heinrich Stolz, Florian Schöne, and Dirk Semkat**
Institut für Physik, Universität Rostock, D-18051 Rostock, Germany
E-Mail: heinrich.stolz@uni-rostock.de

**Abstract:** We demonstrate that the optical linewidth of Rydberg excitons in $Cu_2O$ can be completely explained by scattering with acoustical and optical phonons, whereby the dominant contributions stems from the non-polar optical $\Gamma_3^-$ and $\Gamma_5^-$ modes. The consequences for the observation of polariton effects are discussed. We find that an anti-crossing of photon and exciton dispersions exists only for states with main quantum numbers n>28, so polariton effects do not play any role in the experiments reported up to now.

## 1. Introduction

After the first observation of Rydberg excitons up to main quantum numbers $n$ of 25 in $Cu_2O$ [1], the question arises up to which maximum $n$ Rydberg excitons can be observed in a real experiment [2]. This of course depends primarily on the linewidth, because if it becomes larger than the level distance, the lines start to overlap. Now experimentally it was observed that the linewidth scaled in the same way as the oscillator strength $\sim 1/n^3$, so that a close connection between these two quantities seemed obvious. This opinion was strongly substantiated by the observation that other angular momentum states with less oscillator strength have much smaller linewidth [3]. However, as was already noted some years after the first discovery of the P-exciton series [4] in the 1950s, Toyozawa suggested that the large linewidth of the P-excitons is due to phonon scattering [5]. Furthermore, the strong asymmetry of the lines could be explained by a general theory of phonon-assisted absorption, which take the continuum of the yellow 1S state into account [6]. Recently, the linewidth of the P-exciton absorption lines has been calculated using the full Toyozawa line shape theory [7], but a quantitative explanation of the width and shape of the 2P line was not possible.

The strength of interaction between excitons and photons, usually given as an oscillator strength [8], is important for understanding the basic properties of Rydberg excitons, e.g. the Rydberg blockade. Up to now, these have been discussed with great success in an atomic picture. However, according to a recent publication [9], the P excitons should exhibit strong polariton effects. If this would be true, one has to change the concept of Rydberg excitons completely, as excitons as such do not exist anymore. Instead we have to discuss e.g. dipole-dipole interactions of the polaritons themselves. While in principle the calculation of matrix elements between polariton states is clear (see [10]), the effect of propagation and of the changed vacuum states of the polariton field [10] is completely unknown.

Here, we show that by taking all possible phonon interactions of the P excitons into account, we can explain not only the absolute magnitude of the linewidth, but also the scaling law quantitatively. Furthermore, we discuss the importance of polaritons in the absorption of P excitons and show that only for main quantum numbers larger than 28, which up to know have not been observed, polariton effects will be important, so that the atomic picture survives for excitons which have been investigated up to now.

The paper is organized as follows. In section 2 we derive the scattering matrix elements between arbitrary exciton states for all possible phonon interactions, and discuss their general properties. Furthermore, the kinematics for Stokes and Anti-Stokes scattering for all possible cases, including the continuum states, is deduced. In section 3, the theory is applied to those phonon processes, which contribute most to the linewidth and quantitative scattering rates are derived which are compared to experiment. In section 4, we discuss other scattering processes, like ionization into the continuum. In section 5, we derive the polariton dispersion curves for the multicomponent P excitons and discuss the possibility of a polariton splitting. The results are used to define a correct exciton radiative lifetime.

## 2. Theory of phonon scattering

*2.1 General considerations*

According to the theory of Toyozawa, the linewidth is given by Fermi's golden rule and relates the full width at half maximum of the absorption band to the reciprocal of the total lifetime of the exciton state $\Gamma_i = \hbar / T_{1i}$ [6,11]. $1/T_{1i}$ is given by Fermi's golden rule as

$$1/T_{1i} = \frac{2\pi}{\hbar} \sum_f |M(i \to f)|^2 \delta(E_i - E_f) \ . \tag{2.1}$$

Here $f$ denotes the possible final states of all scattering processes characterized by the transition matrix element M. This first approximation is quite accurate as long as the exciton bands do not overlap and includes also the radiative coupling.

Possible phonon scattering processes for excitons in a perfect crystal (see [12,13]) are:

1. Scattering by acoustical phonons via deformation potentials
2. Scattering by polar optical phonons via Fröhlich interaction
3. Scattering by optical phonons via optical deformation potentials

For two-band Wannier excitons, where electron and hole scatter independently, the matrix elements for all processes have the same form

$$\begin{aligned} M\left(\vec{K}, nlm, \vec{K}', n'l'm', \vec{Q}, \sigma\right) = [&\Xi_c^\sigma(\vec{K} - \vec{K}')W(nlm, n'l'm', \alpha_h \vec{Q}) - \\ &\Xi_v^\sigma(\vec{K} - \vec{K}')W(nlm, n'l'm', -\alpha_e \vec{Q})] \delta(\vec{K} - \vec{K}' \mp \vec{Q}) \end{aligned}, \tag{2.2}$$

with $\Xi_{c,v}^\sigma$ being the general phonon scattering strength of a phonon with $(\sigma, \vec{Q})$. The upper sign denotes Stokes, the lower sign Anti-Stokes scattering. The overlap functions $W$ are given as

$$W(nlm, n'l'm', \vec{Q}) = \int d\vec{r} \phi_{nlm}(\vec{r})^* \phi_{n'l'm'}(\vec{r}) \exp(i\vec{Q} \cdot \vec{r}) \ , \tag{2.3}$$

with $\qquad \alpha_e = m_c^* / M \quad \alpha_h = m_h^* / M \quad M = m_c^* + m_h^* \ . \tag{2.4}$

Since we are interested not only in P-excitons but in general angular momentum states, we cannot choose the coordinate system free as in [7]. Rather, we expand the plane wave in (2.3) into spherical harmonics

$$\exp(i\vec{Q} \cdot \vec{r}) = 4\pi \sum_{l''=0}^{\infty} \sum_{m''=-l''}^{l''} i^{l''} j_{l''}(Qr) Y_{l''m''}^*(\vartheta_Q, \varphi_Q) Y_{l''m''}(\vartheta, \varphi) \tag{2.5}$$

Then the angular integral in (2.3) can be considered as a matrix element of a tensor operator $Y_{l''m''}$ with the angular momentum states $Y_{lm}$ and $Y_{l'm'}$. As the spherical harmonics are an irreducible representation of the spherical group SO(3), the matrix elements are nonzero if and only if the representation $l''$ is contained in the product of the representation of $l$ and $l'$. So $l''$ must be in the range of $|l-l'|$ to $l+l'$ and $m'' = m - m'$. According to the Wigner-Eckart theorem [14] it is given by

$$\langle Y_{lm}|Y_{l''m''}|Y_{l'm'}\rangle = \langle l'm'l''m''|lm\rangle_{CGC} \langle\langle l|l''|l'\rangle\rangle_{red} = GC(l,l',l'',m,m') \; , \tag{2.6}$$

where the first factor denotes the Clebsch-Gordan coefficient and the second denotes the reduced matrix element. Instead of using the Clebsch-Gordan coefficients, for numerical calculations one can use the integrals $GC(l,l',l'',m,m')$ directly.

From inspection of the overlap integrals $W(nlm,n'l'm',\vec{Q})$ we note that all terms in the $l$-expansion have the same $m''$. This means that in calculating the scattering matrix elements we can take out from $W(nlm,n'l'm',\vec{Q})$ a common factor $\exp(im''\varphi_Q)$ which by taking the absolute square drops out. So we can state that the scattering matrix element depends on $|\vec{Q}|$ and the angle of $\vec{Q}$ to the z-axis (denoted by $x = \cos\vartheta_Q$). Note that $W$ in general is either real or imaginary.

Besides the available phase space which is restricted by energy and momentum conservation, the scattering probability between excitons with different quantum numbers $n,l,m$ is determined by the overlap functions. Here we assume that we can approximate the exciton states by hydrogen wave functions (scaled by an appropriate Bohr radius $a_B$). First we notice that scattering within the same exciton states (intraband) always has a term with $l'' = 0$. Since $j_0(0) = 1$, $W(nlm,nlm,\vec{Q})$ always starts at 1 for $Q = 0$ and stays $\approx 1$ up to $Q_{max} = (2l+1)/(n^2 a_B)$. This means that intraband phonon scattering rates at low T and small K are almost independent of n and l, as long as the phonon wave vectors are smaller than $Q_{max}$. For different states (interband scattering) $W(nlm,n'l'm',\vec{Q})$ always starts at 0 for $Q = 0$, so interband scattering requires a finite Q.

As an example of the scaling laws which can be derived from $W(nlm,n'l'm',\vec{Q})$, we discuss the behavior of $W(n10,100,\vec{Q})$ which determines the scattering rate from a P-exciton with main quantum number $n$ to the 1S state. In Fig. 1, we have plotted the dependence of $|W(n10,100,\vec{Q})|$ for different $n$ multiplied by a factor $\sqrt{n^5/(n^2-1)}$. We see that, except for $n=2$, all curves fall on the same dependence. Since the phase space for the scattering which depends on the energy difference between the initial state and the 1S state is almost the same for all $n>2$, we conclude that the linewidth of the P-states scales as the square of $\left|W(n10,100,\vec{Q})\right|^2 \sim (n^2-1)/n^5$ which is just that what is observed experimentally [1]. The maximum of the scaled overlap integral at $Q_{\max} \approx 2/3\, a_B^{-1}$ is the same for all $n$.

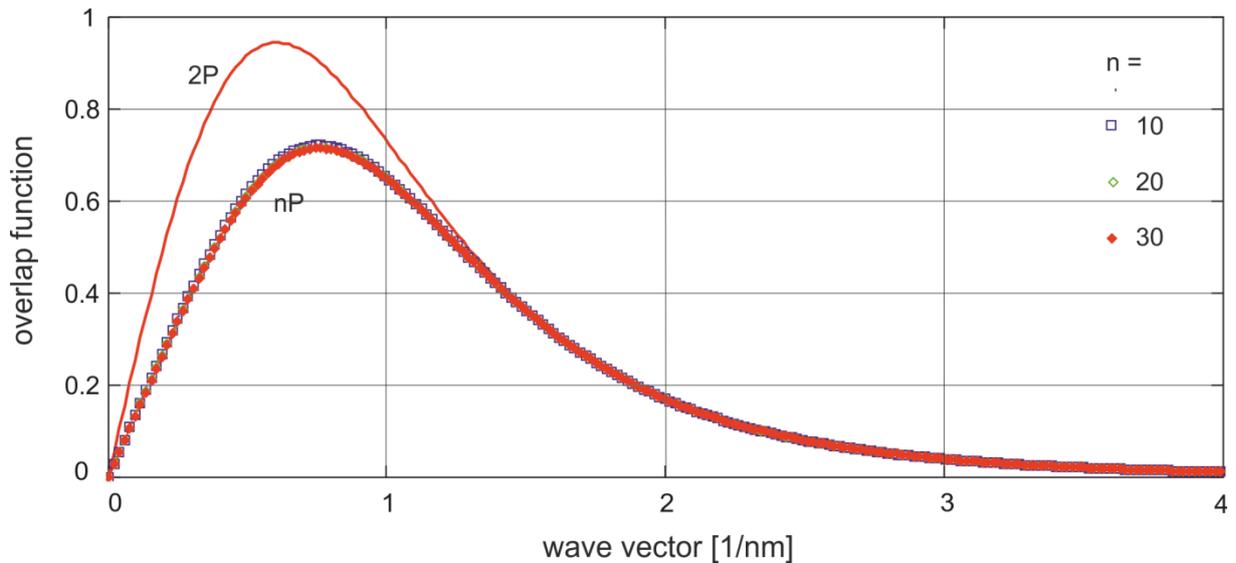

Figure 1: Scaling law for the overlap function between the 1S state and P states with main quantum number $n$ as indicated. Shown are the scaled functions $|W(n10,100,\vec{Q})| n^{5/2}/(n^2-1)^{1/2}$. Red line: $n = 2$; blue, green, and red dotted lines with symbols: $n = 10, 20, 30$.

## 2.2 Interaction strengths

The scattering strengths $\Xi_{c,v}^{\sigma}$ depend on the mechanism and are given for deformation potential scattering by [11,13]

$$\Xi_{c,v}^{LA}(\vec{K}-\vec{K}') = \sqrt{\frac{\hbar}{2\rho N \Omega_0 u_L} \left|\vec{K}-\vec{K}'\right|} \cdot D_{c,v} \, , \qquad (2.7)$$

while for Fröhlich coupling with different LO modes (i=1,2) it is given by [15]

$$\Xi_{c,v}^{LO_i}(\vec{K}-\vec{K}') = \sqrt{\frac{\hbar \omega_{LOi} e_0^2}{2\varepsilon_0 N \Omega_0}} \sqrt{\frac{1}{\varepsilon_i^*}} \cdot \frac{1}{\left|\vec{K}-\vec{K}'\right|} \, . \qquad (2.8)$$

Here $\hbar\omega_{LOi}$ denote the different phonon energies, $\varepsilon_i^*$ denote the effective dielectric constants for the different phonon modes given as

$$\frac{1}{\varepsilon_i^*} = \frac{1}{\varepsilon_i^{up}} - \frac{1}{\varepsilon_i^{low}} \, , \qquad (2.9)$$

where $\varepsilon_i^{up,low}$ denote the dielectric constants above and below the phonon mode which can be calculated from Toyozawa's rule [16]

$$\frac{\varepsilon(\omega)}{\varepsilon_\infty} = \prod_i \frac{(\omega_{LOi}^2 - \omega^2)}{(\omega_{TOi}^2 - \omega^2)} \, . \qquad (2.10)$$

Scattering with the $\Gamma_5^+$ mode is possible for valence band states $\Gamma_8^+$, since $\Gamma_8^+ \otimes \Gamma_5^+ \otimes \Gamma_8^+$ [17] contains the identity representation. This is not the case for the $\Gamma_6^+$ conduction and $\Gamma_7^+$ valence band states, therefore, this phonon process is not important for the yellow exciton states.

Processes involving non-polar optical phonons of symmetry $\Gamma_2^- \oplus \Gamma_3^- \oplus \Gamma_5^-$ are forbidden since neither electron scattering nor hole scattering allow for negative parity phonons.

The odd parity phonons $\Gamma_3^- \oplus \Gamma_5^-$, however, can scatter via a $q$-dependent process since

$$\Gamma_4^- \otimes \Gamma_3^- = \Gamma_4^+ \oplus \Gamma_5^+ \quad \Gamma_4^- \otimes \Gamma_5^- = \Gamma_1^+ \oplus \Gamma_3^+ \oplus \Gamma_4^+ \oplus \Gamma_5^+ \, , \qquad (2.11)$$

and for both $\Gamma_6^+$ and $\Gamma_7^+$ we have $\Gamma_{6,7}^+ \otimes \Gamma_4^+ \otimes \Gamma_{6,7}^+ = \Gamma_1^+ \oplus \Gamma_3^+ \oplus 2\Gamma_4^+ \oplus \Gamma_5^+$ which contains the identity. The first order $\Gamma_2^-$ scattering is forbidden for the yellow states since $\Gamma_4^- \otimes \Gamma_2^- = \Gamma_5^+$.

Information about the magnitude of the corresponding scattering strengths

$$\Xi_{c,v}^{opt,\eta}(\vec{K}-\vec{K}') = \sqrt{\frac{\hbar^2}{2\rho N\Omega_0 E_\eta}} \cdot (\vec{K}-\vec{K}') \cdot \vec{D}_{\eta c,v}^1 \qquad (2.12)$$

with the first order deformation potentials $\vec{D}_{\eta c,v}^1$ can be obtained from the resonant Raman studies in [18], where the observation of both scattering processes has been reported. (see Ref. [18], Fig. 7). Since the cross sections at the same relative energy should depend only on the absolute square of the scattering strengths, we can obtain from the known strength of the $LO_1$ process (given by Eq. (2.8)) the matrix elements of the other processes. At a relative kinetic energy of 13 meV ($\approx 100$ cm$^{-1}$) we have Q=0.9/nm and obtain for $|\vec{D}_3^1| \approx 25 eV$ and $|\vec{D}_5^1| \approx 15 eV$, which are order of magnitude values only. We also cannot distinguish the contributions of valence and conduction band.

Finally, we have to discuss carefully the consequences of the energy conservation in the different type of scattering processes.
The scattering rate between the initial state $|nlm;K>$ and all final states $|n'l'm'>$ is given by

$$\Gamma(nlm,\vec{K};n'l'm') = \frac{2\pi}{\hbar}\frac{N\Omega_0}{8\pi^3}\int_0^\infty dQ\, Q^2 \int_0^\pi \left|M\left(\vec{K},nlm,\vec{K}\pm\vec{Q},n'l'm',Q,\vartheta\right)\right|^2 \delta(E_i-E_f)\sin\vartheta\, d\vartheta \int_0^{2\pi} d\varphi$$
$$(2.13)$$

While the $\varphi$-integration is trivial, the integration over $\vartheta$ (set $x=\cos\vartheta$) requires to determine the zeros of the argument of the delta function $\delta(E_i-E_f) = \delta(f(K,Q,x))$. Here we have to distinguish between 1. Stokes (phonon emission) and 2. Anti-Stokes (phonon absorption) processes. The energies of initial and final states (argument of the delta function in Eq. (2.13)) are

$$\begin{aligned}1)\quad & E_i = E_g - Ry/(n-\delta_{n,l})^2 + \frac{\hbar^2\vec{K}^2}{2M_{Xn}} \quad E_f = E_g - Ry/(n'-\delta_{n',l'})^2 + \frac{\hbar^2(\vec{K}+\vec{Q})^2}{2M_{Xn'}} + \hbar\Omega^\sigma(\vec{Q}) \\ 2)\quad & E_i = E_g - Ry/(n-\delta_{n,l})^2 + \frac{\hbar^2\vec{K}^2}{2M_{Xn}} + \hbar\Omega^\sigma(\vec{Q}) \quad E_f = E_g - Ry/(n'-\delta_{n',l'})^2 + \frac{\hbar^2(\vec{K}-\vec{Q})^2}{2M_{Xn'}}\end{aligned} \qquad (2.14)$$

We see that they depend on the angle $\vartheta$ between $\vec{K}$ and $\vec{Q}$ and on the magnitude of $\vec{Q}$. Note that we are not allowed to set K=0, as the excitons are excited at the finite optical wave vector (see Appendix A). The functions are also slightly different between acoustical and optical phonon scattering. While in the latter processes $\hbar\Omega^\sigma(\vec{Q})$ is approximately constant, in the former the phonon energy scales linearly with Q.

*2.3. Scattering by acoustical Phonons*
For scattering by LA phonons we have

$$f(K,Q,x) = -\frac{Ry}{(n-\delta_{n,l})^2} + \frac{Ry}{(n'-\delta_{n',l'})^2} + \frac{\hbar^2}{2M_X}\beta_{nn'}K^2 - \frac{\hbar^2}{2M_X}(\vec{K}\pm\vec{Q})^2 \mp \hbar u_L Q \qquad (2.15)$$

where we introduced the mass factor $\beta_{nn'} = M_{Xn'}/M_{Xn}$ and set $M_X = M_{Xn'}$ to account for different exciton masses. Actually this is important only for the 1S final state, for all other states $\beta_{nn'} = 1$. Further we define $\kappa = M_X u_L/\hbar$.

If $x_0$ is the zero of $f(K,Q,x)$ as a function of $x$, limited between -1 and +1, the result of the $x$-integration is then simply $1/|f'(x_0)| \left| M\left(\vec{K},nlm,\vec{K}\pm\vec{Q},n'l'm',Q,x_0\right) \right|^2$.

We have to distinguish two situations: Either $\Delta\varepsilon(nl,n'l') = Ry/(n-\delta_{n,l})^2 - Ry/(n'-\delta_{n',l'})^2 \leq 0$ (down scattering, **case a**) or the opposite (up scattering, **case b**).
For **case a)** we define

$$K_a^2 = \left(-\frac{Ry}{(n-\delta_{n,l})^2} + \frac{Ry}{(n'-\delta_{n',l'})^2}\right)\frac{M_X}{\hbar^2}, \qquad (2.16)$$

while for **case b)** we have

$$K_b^2 = \left(\frac{Ry}{(n-\delta_{n,l})^2} - \frac{Ry}{(n'-\delta_{n',l'})^2}\right)\frac{M_X}{\hbar^2}. \qquad (2.17)$$

The integration limits for Q are then obtained as follows:
For the **case 1a)**, the lower limit is

$$Q_1 = -(\kappa+K) + \sqrt{(\kappa+K)^2 + 2K_a^2 + (\beta_{nn'}-1)K^2} \qquad (2.18)$$

while the upper limit is

$$Q_3 = -(\kappa-K) + \sqrt{(\kappa-K)^2 + 2K_a^2 + (\beta_{nn'}-1)K^2}. \qquad (2.19)$$

The case $K_a = 0$ (degenerate intraband scattering) gives $Q_1 = 0$, but $Q_3 = 0$ as long as $K < \kappa$ but $Q_3 = 2(K-\kappa)$ if K is larger. This shows that Stokes scattering is suppressed for excitons at small wave vectors.

In **case 2a)** we obtain as limits for the Q integration

$$\begin{aligned}Q_1 &= (\kappa+K) + \sqrt{(\kappa+K)^2 + 2K_a^2 + (\beta_{nn'}-1)K^2} \\ Q_3 &= (\kappa-K) + \sqrt{(\kappa-K)^2 + 2K_a^2 + (\beta_{nn'}-1)K^2}\end{aligned} \qquad (2.20)$$

In case of $K_0 = 0$ it gives $Q_1 = 2(\kappa+K) \quad Q_3 = 2(\kappa-K)$ which agrees with previous results [19].

For the case of **up scattering (b)** we can set $\beta_{nn'} = 1$ as here only the equal mass case is relevant (the 1S state is so far away from the others that up scattering occurs only at very high temperatures T>300K).

This results **for Stokes scattering (case 1b)** the integration limits are

$$Q_{1,3} = -(\kappa-K) \pm \sqrt{(\kappa-K)^2 - 2K_b^2} \qquad (2.21)$$

This shows that Stokes scattering only occurs if $K \geq \kappa + \sqrt{2}K_b$. In case of **Anti-Stokes scattering (case 2b)** we have as limits for the Q integration

$$Q_{1,2} = (\kappa+K) \pm \sqrt{(\kappa+K)^2 - 2K_b^2}, \quad Q_{3,4} = (\kappa-K) \pm \sqrt{(\kappa-K)^2 - 2K_b^2}. \qquad (2.22)$$

The limits $Q_{1,2}$ are only possible if $K \geq -\kappa + \sqrt{2}K_b$, while $Q_{3,4}$ are possible for $K < \kappa$ and $K_b \leq (\kappa - K)/\sqrt{2}$. If both solutions are possible, one has to integrate from $Q_1$ to $Q_3$ and from $Q_4$ to $Q_2$.

*2.4. Scattering by Optical Phonons*

Due to the low temperature, we have here to consider only case 1a. We assume that the phonon energy $E_{O,i}$ of mode $i$ is independent of Q, so with

$$K_{O,i}^2 = \left( -\frac{Ry}{(n-\delta_{n,l})^2} + \frac{Ry}{(n'-\delta_{n',l'})^2} - E_{O,i} \right) \frac{M_X}{\hbar^2}, \qquad (2.23)$$

the integration limits are

$$Q_1 = -K + \sqrt{2K_{O,i}^2 + \beta_{nn'}K^2}, \quad Q_3 = K + \sqrt{2K_{O,i}^2 + \beta_{nn'}K^2} \qquad (2.24)$$

For the $\Gamma_3^-$ and $\Gamma_5^-$ optical phonons with a first order deformation potential we have the same expressions as for the LO phonons (replace the phonon energy in Eq. (1.47)). Note that in the Q-integral there is an additional factor $Q^2$ due to the matrix element.

### 3. Scattering rates and linewidth

In $Cu_2O$ there is another complication when we consider scattering processes involving the yellow 1S exciton state. Due to the peculiarities of the valence band structure, it has a much larger binding energy (ortho 1S 140 meV compared to the exciton Rydberg of 87.653 meV) concomitant with a smaller exciton radius of only half of the Bohr radius ($a_B = 1.1$ nm). Furthermore, the effective mass is also much larger than the sum of electron and hole masses. We will take these effects into account by using in the overlap integral the smaller Bohr radius and in the energy conservation the larger exciton mass, but neglect the effect on $\alpha_e$ and $\alpha_h$.

The interaction strength of LA scattering depends on the deformation potentials. In Ref. [7] the authors used $D_e = 2.4$ eV and $D_h = 2.2$ eV taken from Ref. [20]. However, we think that these values are highly questionable.

1. For scattering of 1S paraexcitons where $W = 1$ for both electrons and holes at exciton wave vectors smaller than $1/a_B$ the value $D_e - D_h = 1.7$ eV is very well established from diffusion experiments [21], strain measurements [22] and paraexcitons propagation beats [23]. Obviously, assuming $D_e - D_h = 0.2$ eV contradicts these results.
2. Measurements of Hall mobility gave $D_h = 0.7$ eV [24], which would fit exactly.
3. However, a closer inspection of the theory used in Ref. 20 shows, that their interpretation of the experimental results is incorrect. To derive the deformation potentials from the measured T-dependence of the cyclotron resonance linewidth, they use a theory which assumes sufficiently high temperature, so that equipartition of the phonon modes occurs (see [12]). This is not valid for $Cu_2O$ due to the high masses compared to usual semiconductors. Therefore, one has to use the exact expression for the scattering rate at low temperatures (see [25]). A re-evaluation of their results with the correct theory gives as deformation potentials $D_e = 3.5$ eV and $D_h = 1.8$ eV, the difference of which would exactly give the right exciton 1S deformation potential of 1.7 eV.

In our calculations we therefore use for the deformation potentials $D_e = 3.5$ eV and $D_h = 1.8$ eV. For the sound velocity we use $u_L = 4500$ m/s.

For multimode LO scattering the phonon energies are [26]

$$E_{TO1} = 18.8 meV \quad E_{LO1} = 19.1 meV$$
$$E_{TO2} = 78.5 meV \quad E_{LO2} = 82.1 meV \quad (2.25)$$

from which we have to determine the effective $\varepsilon_i^*$. We use

$$\varepsilon_{stat} = 7.37 \quad \varepsilon_{int} = 7.14 \quad \varepsilon_\infty = 6.53 \quad (2.26)$$

resulting in $\varepsilon_1^* = 233$ $\varepsilon_2^* = 76$. As electron and hole mass we use the polaron masses $m_c^* = 0.985 m_e$ $m_v^* = 0.575 m_e$ [25], which gives as exciton masses for the states with $n \geq 2$ $M_X = 1.56 m_e$. For the 1S state, the exciton mass is different due to the valence band structure and is given by $M_X = 2.61 m_e$ [27] (note that we use the same value for the ortho- and paraexciton mass, neglecting the anisotropy of the orthoexciton mass due to the complex valence band structure [28]). To make the calculations more simple, we further combine the $\Gamma_3^-$ and $\Gamma_5^-$ processes in one with an effective D of 38 eV.

*3.1 nP states*

Here we first show the results for the contributions of the different scattering processes for the nP

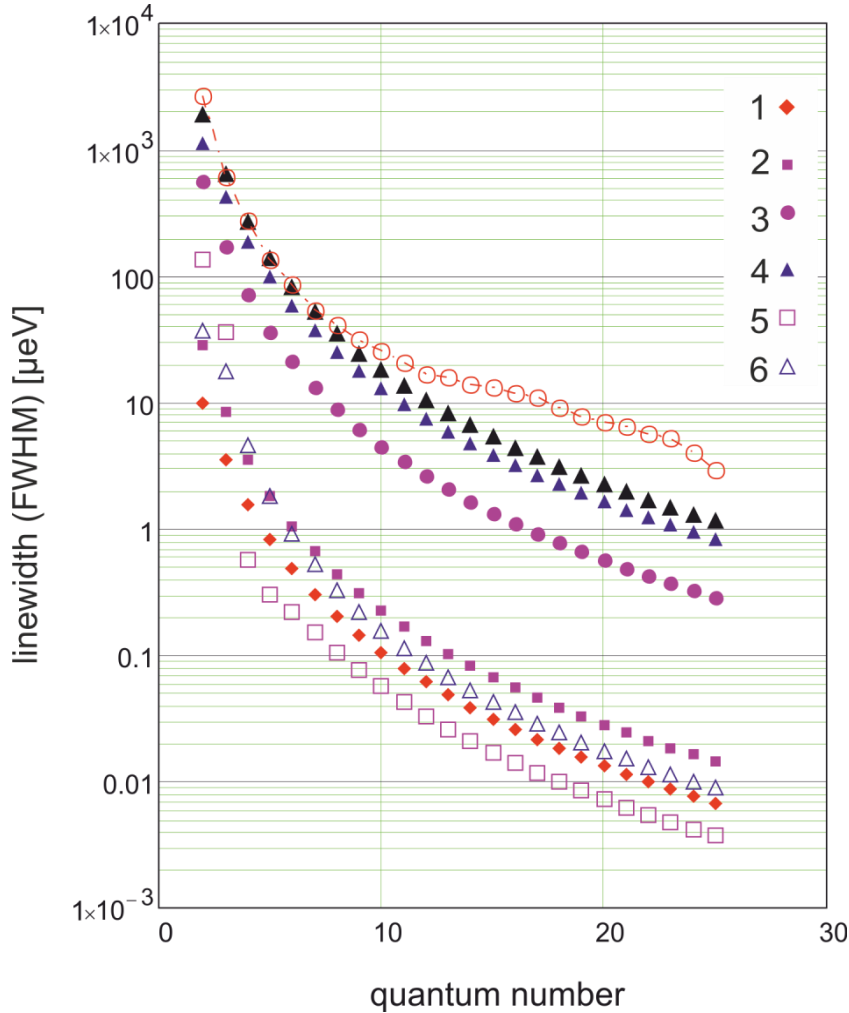

Figure 2:
Contributions to the linewidth of the nP states (in μeV) from different scattering processes.
1: acoustic deformation potential,
2: LO1 Fröhlich scattering,
3: LO3 Fröhlich scattering,
4: $\Gamma_3^- / \Gamma_5^-$ deformation potential,
5: LO1 to the 2S state,
6: $\Gamma_3^- / \Gamma_5^-$ to the 2S state.

The red circles are the experimentally deduced linewidths from [1].

-8-

exciton states. In Fig. 2, the partial linewidths due to scattering by the LA mode, the two LO modes and the $\Gamma_3^- / \Gamma_5^-$ modes are shown. The total linewidth obtained by summing all contributions agrees quite well with the experimental data (red circles) taken from Ref. [1] up to n=9. Even the value for the 2P state differs from experiment only by 25%. We expect this small difference to vanish if we would use the full Toyozawa theory [6] which should have the strongest effect for the 2P state. Obviously, the dominant contribution comes from the LO2 and the $\Gamma_3^- / \Gamma_5^-$ processes, whereby the latter contributes about 2/3 to the total linewidth. In all previous calculations, it was just this process which was neglected explaining their failure.

Since the optical phonon energies involved in the scattering are all above 10 meV, the temperature dependence of the scattering rates which is given approximately by $1 + 2n_B(\hbar\Omega)$ is negligible up to 50 K. The acoustical modes which would already increase strongly above 10 K, contribute less than 1% to the linewidth, so their influence can be neglected. Up to 50 K we, therefore, predict the linewidth to be temperature independent.

In conclusion, we can explain the linewidth of al P exciton states by the following scattering processes:

1. Scattering by LO1 and LO2 phonons via the Fröhlich mechanism
2. Scattering by $\Gamma_{3,5}^-$ phonons by a first order deformation potential scattering.

## 4. Thermal ionization of Rydberg excitons by phonons

A special case of up scattering is the thermal ionization of excitons by Anti-Stokes LA scattering of states with high quantum number into the continuum. To include in the theory a possible electron-hole plasma interacting with the excitons [29], we include a band gap shift of $-\Delta$ in the calculation. A closer look on Eq. (2.22) shows, that the thermal ionization (for $\Delta = 0$) sets in, when $K_b \leq (\kappa + K_{opt})/\sqrt{2}$ which happens for n>21. The second range of wave vectors (Q3,Q4) opens up if $K_b \leq (\kappa - K_{opt})/\sqrt{2}$ which requires n>70. Note that, since $\kappa > K_{opt}$, Stokes scattering is not possible. $K_b$ is now given by

$$K_b^2 = \left( \frac{Ry}{(n-\delta_{n,l})^2} - \Delta \right) \frac{M_X}{\hbar^2} \ . \tag{4.1}$$

Therefore, the range of possible $n$ changes with increasing $\Delta$ to smaller values.

In the continuum the exciton states are determined besides the angular momentum $l,m$ by a continuous quantum number $k$ and by the center of mass momentum $\hbar\vec{K}$. The energy is given by

$$E(k,l,m,\vec{K}) = E_g + \frac{\hbar^2 k^2}{2\mu_{eh}} + \frac{\hbar^2 \vec{K}^2}{2M_{eh}} \tag{4.2}$$

with $1/\mu_{eh} = 1/m_c + 1/m_h$ and $M_{eh} = m_c + m_h$.

The scattering rate is now given by the expression

$$\Gamma(nlm,\vec{K}) = \frac{2\pi}{\hbar} \frac{N\Omega_0}{8\pi^3} \frac{N\Omega_0}{8\pi^3} \int d^3Q \sum_{l'm'} \int dk \left| M\left(\vec{K},nlm,\vec{K} \pm \vec{Q},k,l'm',Q,\vartheta\right) \right|^2 \delta\left(E_i - E_f\right). \quad (4.3)$$

The matrix elements have to be calculated with the continuum wave functions

$$\varphi_{k,l,m}(\vec{r}) = Y_{l,m}(\vartheta,\phi) R_{k,l}(r)$$
$$R_{k,l}(r) = \frac{1}{\sqrt{N\Omega_0}} e^{\frac{1}{2}\pi\alpha} \left|\Gamma(l+1+i\alpha)\right| (2kr)^l \cdot e^{-ikr} F(i\alpha+l+1;2l+2;2ikr)/(2l+1)! \quad (4.4)$$

with $F$ being the confluent hypergeometric function and

$$\alpha = \frac{\sqrt{2\mu_{eh}Ry}}{\hbar k} = \frac{1}{a_B k}. \quad (4.5)$$

In a first approximation, we will use plane waves with wave vector $\vec{k}$ and energy $\hbar^2 k^2 / 2\mu_{eh}$ for the relative motion and according to Toyozawa [30] multiply the matrix elements by the square root of the Sommerfeld factor $2\pi/(a_B k)$. The integration in (4.3) then has to be done over the $\vec{k}$ space.

Expressing the Fourier transform as a convolution one can easily calculate the overlap functions of an $n,l,m$ state with a continuum state as

$$W(nlm,\vec{k},\vec{Q}) = \int d\vec{r}\, \phi_{nlm}(\vec{r})^* \frac{1}{\sqrt{N\Omega_0}} \exp(i\vec{k}\cdot\vec{r}) \exp(i\vec{Q}\cdot\vec{r}) = $$
$$= \frac{1}{\sqrt{N\Omega_0}} \tilde{\phi}_{nlm}(\vec{k}+\vec{Q})^* = \frac{1}{\sqrt{N\Omega_0}} \tilde{R}_{nl}(|\vec{k}+\vec{Q}|) Y^*_{lm}(\vartheta_{\vec{k}+\vec{Q}},\varphi_{\vec{k}+\vec{Q}}) \quad (4.6)$$

with $\tilde{R}_{nl}(k)Y_{lm}(\vartheta,\varphi)$ being the Fourier transform of the hydrogen wave function and

$$\tilde{R}_{nl}(k) = 2^{2l+4}(na_B k)^l \frac{l!}{(1+(na_B k)^2)^{l+2}} (na_B/2)^{3/2} \cdot \sqrt{\frac{n(n-l-1)!}{\pi(l+n)!}} \cdot GegC\left[l+1, n-l-1, \frac{(na_B k)^2 - 1}{(na_B k)^2 + 1}\right]$$
$$(4.7)$$

$GegC$ are the Gegenbauer polynomials [31].

In the following we only consider the case of Anti-Stokes LA scattering (due to the low temperature optical phonon modes are not occupied) and restrict to the case with 21<n<70.

We start off with the integration over the angles of Q. While the $\varphi_Q$ integration is trivial, we utilize the delta-function for the $\vartheta_Q$ integration. Denoting the zero as $x_{0Q} = \cos(\vartheta_{0Q})$ we have to multiply the integrand by $1/|f'(x_{0Q})|$ and insert $\vartheta_Q = \vartheta_{0Q}$. As limits for the Q integration we get

$$Q_{\min} = (\kappa + K_{opt}) - \sqrt{(\kappa + K_{opt})^2 - 2K_b^2} \quad \text{and} \quad Q_{\max} = (\kappa + K_{opt}) + \sqrt{(\kappa + K_{opt})^2 - 2K_b^2}.$$

$k_0(Q)$ determines the upper limit of the k-integration and is given by

$$k_0(Q) = \sqrt{\frac{\mu_{eh}}{M_{eh}}} \left(2(\kappa + K_{opt})Q - 2K_0^2 - Q^2\right)^{1/2}. \quad (4.8)$$

The integrand depends, however, on the angle between $\vec{k}$ and $\vec{Q}$. But for the $\varphi_k$ and $\vartheta_k$ integration we can rotate the coordinate system in such a way that $\vec{k} \| \vec{Q}$. Then the integrand depends only on the new $\vartheta_k$, but not on $\varphi_k$. Then the second integration is trivial and we have only to perform the first one giving for the scattering rate ($x = \cos\vartheta_k$):

$$\Gamma_{ion}(nlm,\vec{K}) = \frac{2\pi}{\hbar}\frac{2\pi}{8\pi^3}\frac{2\pi}{8\pi^3}\frac{\hbar}{2u_{LA}\rho_0}\frac{M_{eh}}{\hbar^2}\frac{1}{K}\int_{Q_{min}}^{Q_{max}} Q^2 n_B(\hbar u_{LA}Q)dQ \int_0^{k_0(Q)} 2\pi/(a_Bk)k^2 dk \int_{-1}^{+1} dx \cdot$$
$$\left[\left[D_c\tilde{R}_{nl}\left(\sqrt{\sigma_h^2 Q^2 + k^2 + 2\sigma_h Qkx}\right)Y_{lm}(x) - D_v\tilde{R}_{nl}\left(\sqrt{\sigma_e^2 Q^2 + k^2 - 2\sigma_e Qkx}\right)Y_{lm}(-x)\right]\right]^2$$

(4.9)

with $n_B(E)$ being the Bose function (Anti-Stokes scattering!).

The results of the calculation for $\Delta = 0$ are shown in Fig. 3.

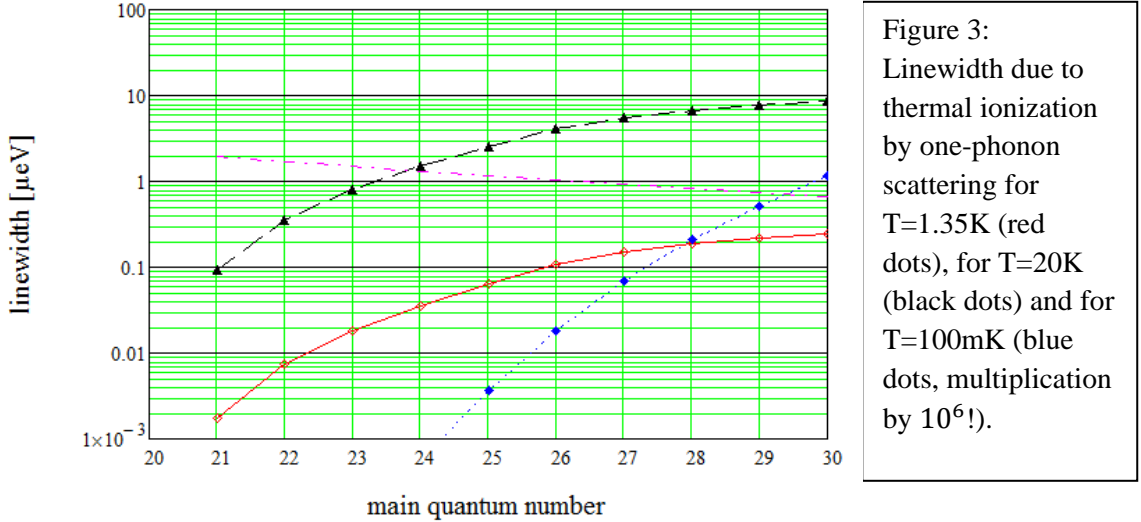

Figure 3: Linewidth due to thermal ionization by one-phonon scattering for T=1.35K (red dots), for T=20K (black dots) and for T=100mK (blue dots, multiplication by $10^6$!).

The red curve for T=1.35K shows that ionization is **not** important for the observation limits for higher n states, as the rate becomes larger than the lifetime only for n > 30. For 20K, thermal ionization sets in for n > 24, for very low T=100mK it is completely negligible.

## 5. Oscillator strength, radiative lifetime and polariton effect for the P states in $Cu_2O$

The question, whether the P excitons do form polaritons is of central importance for the properties of Rydberg states. In our discussion we start from the classical exciton-polariton (see e.g. [32]).

Consider first a single resonant exciton state with energy

$$E_X(K) = E_0 + \frac{\hbar^2}{2M}K^2 - i\gamma(K) \quad (5.1)$$

where M is the mass and $\gamma(K) = \hbar/2T_1(K)$ is a homogeneous broadening, $T_1(K)$ being the total lifetime of the state (see Eq. (2.1)).

-11-

The state has an optical transition, which we characterize by an oscillator strength density $f/V$ (V is the crystal volume) which is related for a forbidden (second class) transition to the band-to-band dipole $d_{cv} = \langle v|e\vec{r}|c\rangle$ by the relation [8,32]

$$f/V = \frac{4\pi\varepsilon_0}{e_0^2}\frac{m_0 E_X}{\hbar^2} 2g_O |d_{cv}|^2 \cdot \left|\nabla\varphi_X(\vec{r})\right|_{\vec{r}=0}\right|^2 . \tag{5.2}$$

Here $\varphi_X(\vec{r})$ is the exciton envelope function (usually assumed as hydrogen-like) and $g_O = 4/3$ is a degeneracy factor taking the singlet part of the ortho states into account. For second class transitions the dipole moment may depend on the direction of the light propagation (like for a quadrupole transition).

Note that for excitons the oscillator strength itself (and all derived quantities like transition dipole moment and radiative lifetime) is meaningless since it is proportional to the crystal volume and thus diverges in the limit $V \to \infty$.

The contribution of the exciton state to the dielectric function is given by the dispersion relation

$$\varepsilon(\vec{k},\omega) = \varepsilon_\infty + \frac{4\pi\beta E_X(\vec{k})^2}{E_X(\vec{k})^2 - (\hbar\omega)^2} . \tag{5.3}$$

The strength constant $\beta$ is related to the oscillator strength density by

$$\beta = \frac{\hbar^2 e_0^2}{4\pi\varepsilon_0 m_0 E_X^2}\frac{f}{V} . \tag{5.4}$$

The (transverse) polaritons are then solutions of the equation

$$\frac{c_0^2 k^2}{\omega^2} = \varepsilon(k,\omega) . \tag{5.5}$$

The splitting between transversal and longitudinal excitons (LT-splitting) is directly related to the oscillator strength density and to the strength parameter $\beta$ by

$$\Delta E_{LT} = \frac{\hbar^2 e_0^2}{2\varepsilon_0 \varepsilon_\infty m_0 E_X}\frac{f}{V} = \frac{2\pi\beta}{\varepsilon_\infty}E_X(0) . \tag{5.6}$$

To determine the polariton dispersion one only needs to know $E_X, \beta, \varepsilon_\infty$. Both dipole moment and oscillator strength are redundant quantities.

One can simplify the dispersion relation by applying the rotating wave approximation, i.e., assuming $\hbar\omega \approx E_X$. Then we can write $E_X^2 - (\hbar\omega)^2 = 2E_X(E_X - \hbar\omega)$ and get as dispersion relation

$$\varepsilon(\vec{k},\omega) = \varepsilon_\infty + \frac{2\pi\beta E_X(\vec{k})}{E_X(\vec{k}) - (\hbar\omega)} . \tag{5.7}$$

As the higher P lines are lying very close together, one has to take their total contribution to the dielectric function into account. Generalizing (5.7) to the case of many resonances, we have

$$\frac{c_0^2 k^2}{\omega^2} = \varepsilon(\vec{k}, \omega) = \varepsilon_\infty + \sum_{i=1}^{N} \frac{2\pi\beta E_{X,i}(\vec{k})}{E_{X,i}(\vec{k}) - (\hbar\omega)} \ . \tag{5.8}$$

To calculate the polariton dispersion relations for $Cu_2O$ we have to know the strength parameters, the energies of the P states and their damping.

The oscillator strength per unit cell for the $n = 2$ P line was given in Ref. [33] as $2.7 \cdot 10^{-6}$. Using this value and the dependence of the oscillator strength on $n$ which is given by $(n^2 - 1)/n^5$, we can calculate for each n the oscillator strength density as

$$f(n)/V = 2.88 \cdot 10^{-5} \cdot \frac{n^2 - 1}{n^5} \cdot \frac{1}{a_L^3} \ , \tag{5.9}$$

which gives as strength parameters

$$\beta(n) = 4.1107192 \cdot 10^{-5} \frac{eV^2}{E_P(n)^2} \cdot \frac{n^2 - 1}{n^5} \ . \tag{5.10}$$

The energies of the nP states (important are those for n>10) have been obtained by fitting the measured values (from [1]) with a quantum defect formula [34]

$$E_P(n) = E_g - \frac{Ry}{(n - \delta_P)^2} \tag{5.11}$$

with

$$\begin{aligned} E_g &= 2172.0483 \text{ meV} \\ Ry &= 87.653 \text{ meV} \qquad , \\ \delta_P &= 0.1986 \end{aligned} \tag{5.12}$$

which allow to reproduce the measured energies with an error of less than $\pm 0.5\,\mu eV$. Note that the value of the band gap differs from that in [1] as one has to subtract from the resonance energies the energy of the excitons at the crossing point with the light dispersion E=26.96 µeV (which is almost constant for n>10).

As we have shown in Section 2, the damping of the polaritons originates from phonon scattering (for the influence of radiative coupling see below). One can use for our purpose the approximate formula

$$\hbar\gamma(n) = \frac{n^2 - 1}{n^5} \cdot 9 \ meV \ . \tag{5.13}$$

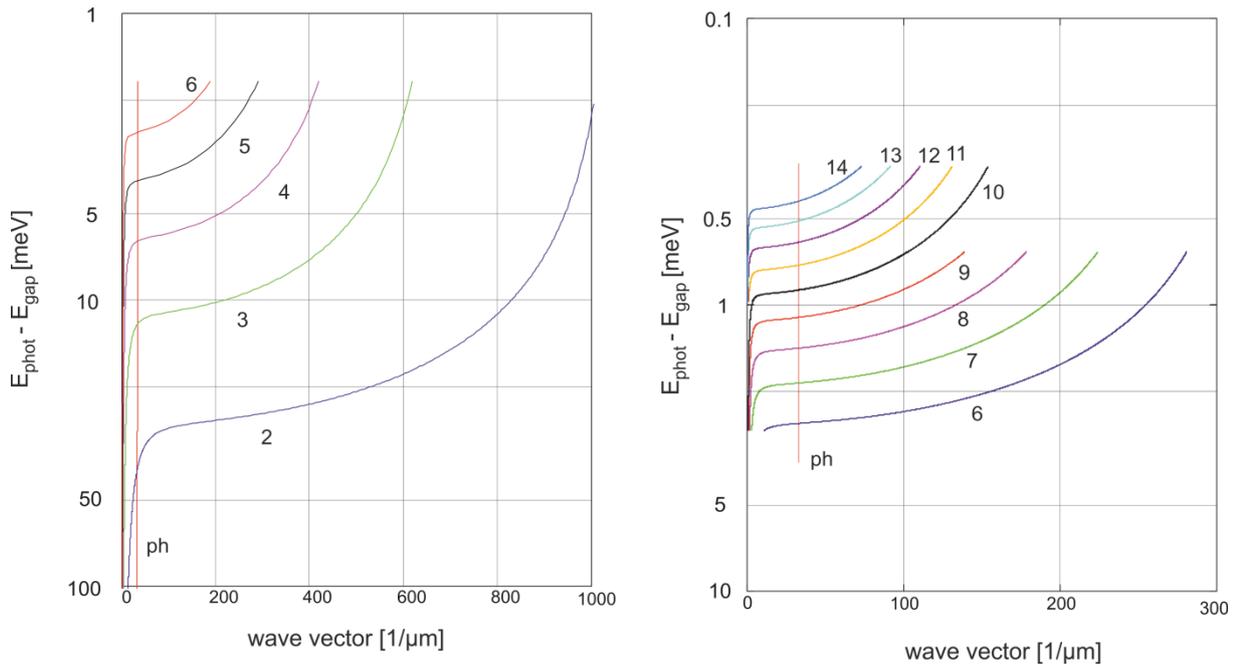

Figure 4: The dispersion relations for the P polariton states from n=2 to n=14. ph denotes the dispersion of the photon mode.

Now we are able to calculate the polariton dispersions for all Rydberg states. For the calculation we use the method of Cho [35]. The results are shown in Fig. 4 for the real part and Fig. 5 for the imaginary part.

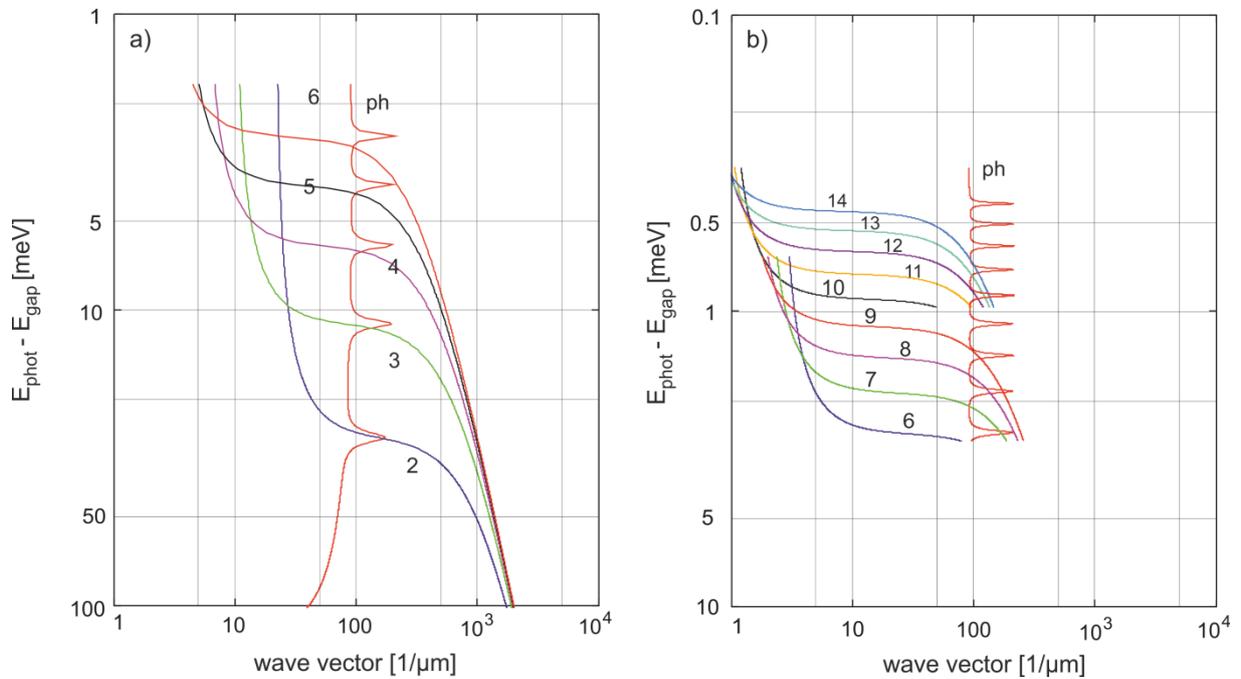

Figure 5: Imaginary part of the polariton wave vector for the P polariton states n=2 to 6 (part a) and n=6 to n=14 (part b).

Obviously, the photon line crosses all exciton dispersions without any disturbance: this means that the polariton character as a propagating wave is completely absent. The reason for this can be seen in the imaginary part of the wave vector, which is the inverse damping length of the polariton wave. In the resonance region (where the photon damping has its maximum and directly gives the absorption coefficient (multiplied by $2 \cdot 10^4$ to give the value in 1/cm, the peak heights are fully in agreement with experiment), the damping of the polariton waves corresponding to the excitons is about 100/µm. This means that these waves are damped out within 10 nm! Consequently we have no polariton splitting by avoided crossing.

This picture is completely different from that in Ref. [8], where typical polariton dispersion relations for the P excitons have been claimed. These results have to be considered doubtfull for several reasons: One is that the strength of the light-matter coupling is strongly overestimated. This is related to the wrong value of the LT splitting that is used by the authors. Instead of their value 10 µeV it is actually only 1.25 µeV for the n=2 P state. Second, the exciton mass used by these authors is only about 0.015 $m_0$, compared to the real mass of about 1.6$m_0$. Third, unfortunately the authors do not specify the damping constants, but they seem to be much too small.

Still the question remains, whether a polariton splitting occurs for larger quantum numbers. Detailed calculations, which are reproduced in Fig. 6 show that for n=30 and beyond an extremely small polariton splitting is visible. Obviously, any experimental verification of these splittings by e.g. polariton beats [23,35] will be a challenging task!

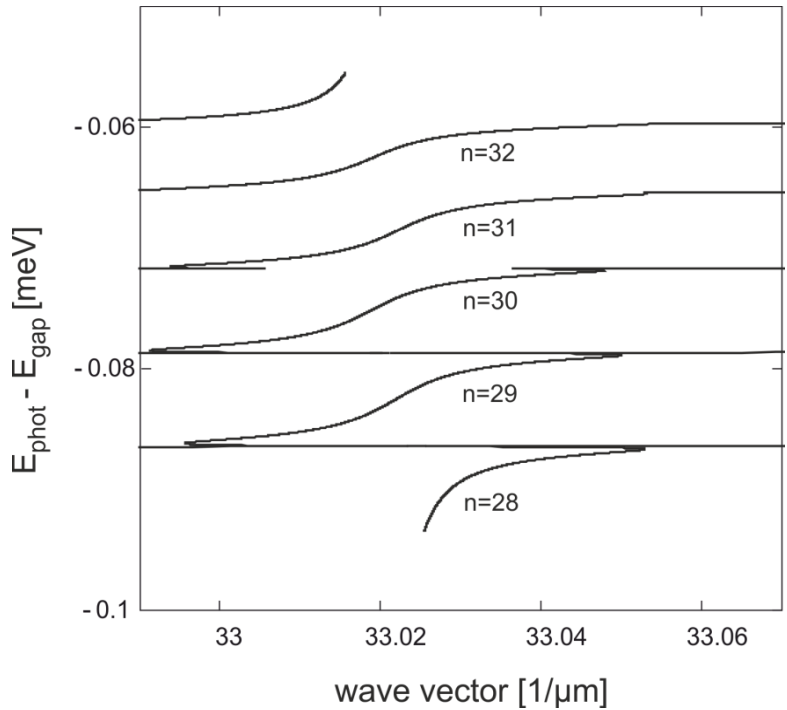

Figure 6: Polariton dispersion relations for n=28 to 32. For $n \geq 30$ we observe a clear splitting of the dispersion relations, which, however, is extremely small ($\Delta k \approx 0.02$ µm$^{-1}$ and $\Delta E \approx 10$ neV).

Despite the lack of anti-crossing, the polariton concept has important consequences for the exciton dynamics, as it allows one to define unambiguously the *coherence volume* of the excitons [32]. As one sees from Eq. (5.2), the oscillator strength depends on the volume of the sample, which taking literally, would make the concepts of atomic physics obsolete. This, however, requires that the exciton translational motion is coherent over the total crystal volume. As the excitons are scattered by phonons, they lose their coherence over the mean free path due to the scattering processes, which is given by the ratio of group velocity over scattering rate. In the polariton picture, one should use for consistency the polariton group velocity $V_{gr}$, which can be determined from the dispersion curves at the resonance frequency. The values obtained from the above dispersion curves are shown in Fig. 7 (red diamonds). While it decreases with n for n<8, it stays constant above. Therefore, the mean free path increases with n according to the decrease of the lifetime.

$$L_{coh}(n) = V_{gr}(n) / \gamma_{tot}(n) \qquad (5.14)$$

As usual, $\gamma_{tot} = \gamma_{rad} + \gamma_{nrad}$ whereby $\gamma_{nrad} = \hbar / 2T_1$ is the non-radiative dephasing rate due to e.g. phonon scattering. Assuming that the excitons within a sphere of radius $L_{coh}/2$ emit coherently, the

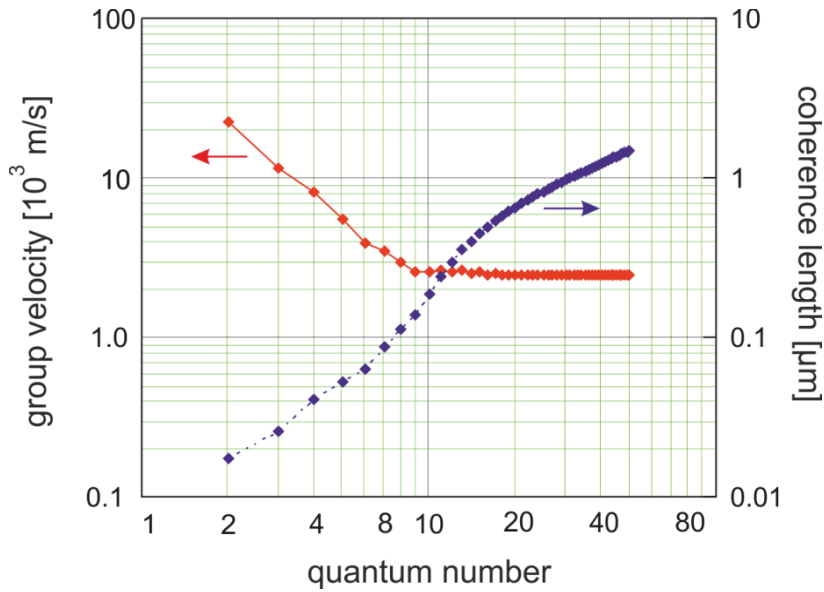

Figure 7: Polariton group velocities (red diamonds) and coherence length (blue diamonds) of the P excitons for n=2 to 40. Up to n=9 the coherence length is determined solely by the group velocity and non-radiative damping, for higher n by both radiative and non-radiative decay.

radiative lifetime is given by

$$\gamma_{rad} = \frac{n_b \hbar e_0^2}{24 \varepsilon_0 m_0 c_0^3} \left(\frac{E_X}{\hbar}\right)^3 \left(\frac{L_{coh}}{a_L}\right)^3 \cdot f/V \qquad (5.15)$$

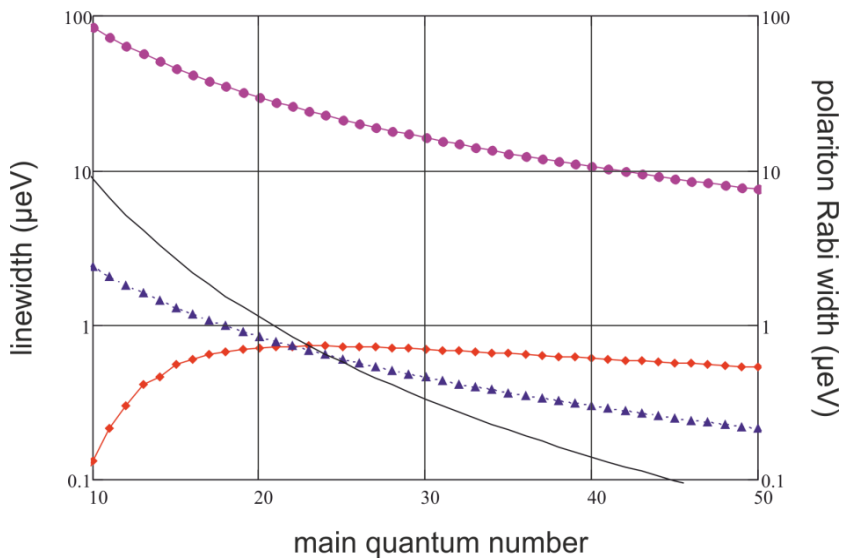

Figure 8: Radiative broadening for P-excitons for main quantum numbers from 10 to 50 (red diamonds) after Eq. (5.15). The magenta dots represent the polariton Rabi frequencies (Eq. (5.16)) and the black full line the nonradiative widths due to phonon scattering (Eq. **Anzuzeigender Text darf nicht m** . The blue triangles show the n dependence of the spatial coherence critical linewidth

These two equations have to be solved self-consistently. The results for the coherence length are shown in Fig. 7 (blue diamonds), for the radiative rate in Fig. 8 (red diamonds). We see that the radiative broadening never gets larger than 1 µeV, but for n>26 determines the width of the resonances. So we expect polariton effects to become important. Indeed the dispersion relations for n>28 do show a (very) small polariton splitting (see Fig. 6).

Actually, our results agree with the well-known criteria [32,36] for the existence of the polariton effect. Here we have to distinguish between two situations: (i) the quasi-particle picture, where one creates polaritons with well-defined wave vector and (ii) the forced harmonic situation where polaritons with well-defined frequency are created by an external harmonic driving source, e.g., an electromagnetic wave. The latter is appropriate for transmission experiments, while the first applies, e.g., to two-photon absorption.

The existence of polaritons in the first case is governed by the condition that the (polariton) Rabi frequency

$$\Omega_R = \sqrt{\Delta E_{LT} \cdot E_X(0)} / \hbar \qquad (5.16)$$

is larger than the damping $\gamma$ (temporal coherence). In the second case, polaritons exists (in the sense that one can observe a non-crossing of the dispersion relations) if the following criterion is fulfilled (spatial coherence)

$$\gamma < \gamma_c = \sqrt{\frac{8\Delta E_{LT}}{\hbar^2 M_X c^2 / n_b^2} \cdot E_X(0)^2} \qquad (5.17)$$

A plot of these quantities (Fig. 8) shows that forced harmonic polaritons would exist for nP>28, which is in nice agreement with our direct calculation of the dispersion curves. In contrast, quasi-particle polaritons would already show up for P excitons with quantum number n>3, which has important consequences for the excitation of these states via a two-step process involving the yellow orthoexciton and a suitable mid-infrared laser supplying the energy of the 1S to nP transition [38]. In such an experiment we expect not only the observation of pronounced interference fringes similar to those in two-photon excitation of the blue exciton states in $Cu_2O$ [39] but complete new aspects in the physics of Rydberg excitons, like a polariton-polariton blockade.

## 6. Conclusions

We have presented detailed calculations for the total linewidth of exciton states in $Cu_2O$ by considering the interactions with phonons and photons. Taking not only the well-known scattering with acoustical and polar optical phonons but also with non-polar optical phonons into account, we are able to deduce the linewidth of the P exciton states with angular momentum $l = 1$ in almost quantitative agreement with experiments. We further show that for main quantum numbers n<28 the polariton effect does not lead to a splitting of the dispersion relations, in contrast to previous studies [8]. We further exploit the concept of exciton coherence length to obtain the radiative linewidth of the excitons. Only for n>28 it becomes larger than the linewidth due to phonon scattering, so that for these, up to now not observed, exciton states the radiative coupling, i.e., the polariton character, dominates. Our results should clarify the roles of phonon and photon coupling for Rydberg excitons and open the way to more advanced experiments.

**Acknowledgements:** We would like to thank M. Bayer (Dortmund) for fruitful discussions and the Collaborative Research Center (SFB) 625 funded by the Deutsche Forschungsgemeinschaft for financial support of this work.

**Appendix A: Determination of the wave vector of the optically excited exciton states**

The wave vector of the optically excited excitons is given as the solution of the equation

$$E_g - Ry/n^2 + \frac{\hbar^2}{2M_X}K_{opt}^2 = \hbar \frac{c_0}{n_b} K_{opt} \tag{A.1}$$

This quadratic equation has the solutions

$$K_{1,2} = \frac{M_X c_0/n_b}{\hbar} \pm \sqrt{\left(\frac{M_X c_0/n_b}{\hbar}\right)^2 - \frac{2M_X(E_g - Ry/n^2)}{\hbar^2}} \tag{A.2}$$

Since the last term in the root is much smaller than the first, we can expand the root and get with an error of $10^{-5}$

$$K_{opt} = \frac{(E_g - Ry/n^2)}{\hbar c_0/n_b} \tag{A.3}$$

with $E_g$ = 2.17208 eV and Ry=87.653 meV, we get for n=20 $K_{opt} = 3.302 \cdot 10^7 m^{-1}$.

**Appendix B: Properties of some overlap integrals**

As examples we first discuss the case of S → S scattering. Here $l = 0, l' = 0$ and there is only one contribution $l'' = 0$. This is given by

$$W(n00,n'00,\vec{Q}) = 4\pi Y_{00}^*(\vartheta_Q,\varphi_Q) CG(1,0,0,0,0) \int r^2 dr R_{n0}(r)^* R_{n'0}(r) j_0(Qr) . \tag{A.4}$$

For P → S scattering we have $l = 1$ and $l' = 0$, so that $l'' = 1$. For $m = 0, m'' = 0$ and we have

$$W(n10,n'00,\vec{Q}) = 4\pi i Y_{10}^*(\vartheta_Q,\varphi_Q) CG(1,0,1,0,0) \int r^2 dr R_{n1}(r)^* R_{n'0}(r) j_1(Qr) \tag{A.5}$$

For $m = 1$ we get

$$W(n11,n'00,\vec{Q}) = 4\pi i Y_{11}^*(\vartheta_Q,\varphi_Q) GC(1,0,1,1,0) \int r^2 dr R_{n1}(r)^* R_{n'0}(r) j_1(Qr) \tag{A.6}$$

Both expressions, except for a factor $e^{i\varphi_Q}$ which cancels by taking the absolute value, are the same showing the spherical symmetry.

On the other hand, for P → P scattering, we get 3 terms since $l'' = 0, 1$, and 2. But because $CG(1,1,1,m,m') = 0$ only two remain

$$\begin{aligned}W(n10,n'10,\vec{Q}) = &4\pi Y_{00}^*(\vartheta_Q,\varphi_Q) CG(1,1,0,0,0) \int r^2 dr R_{n1}(r)^* R_{n'1}(r) j_0(Qr) - \\ &- 4\pi Y_{20}^*(\vartheta_Q,\varphi_Q) CG(1,1,2,0,0) \int r^2 dr R_{n1}(r)^* R_{n'1}(r) j_2(Qr)\end{aligned} \tag{A.7}$$

For the general case of P->L scattering we have also in total two terms since $l'' = L-1, L, L+1$, but $CG(1, L, L, m, m') = 0$.